# Quantum-Assisted Tomographic Image Refinement with Limited Qubits for High-Resolution Imaging

Hyunju Lee and Kyungtaek Jun

***Abstract*—We propose a quantum-assisted reconstruction framework for high-resolution tomographic imaging that significantly reduces both qubit requirements and radiation exposure. Conventional quantum reconstruction methods require solving QUBO (Quadratic Unconstrained Binary Optimization) problems over full-resolution image grids, which limits scalability under current hardware constraints. Our method addresses this by combining sinogram downscaling with region-wise iterative refinement, allowing reconstruction to begin from a reduced-resolution sinogram and image, then progressively upscaled and optimized region by region.

Each region is independently transformed into a compact QUBO problem and solved via a D-Wave hybrid quantum-classical solver while keeping the surrounding image fixed. The framework supports both full-view and sparse-view sinograms and is compatible with diverse acquisition geometries, including parallel-beam, fan-beam, cone-beam, synchrotron tomography, and electron tomography (ET). Adaptive and overlapping region selection is also supported to enable targeted refinement of structurally important areas.

Experimental validation on binary and integer-valued Shepp-Logan phantoms demonstrates accurate reconstructions under both dense and sparsely sampled projection conditions using significantly fewer qubits. We observed that nearest-neighbor interpolation may cause edge artifacts that hinder convergence, which can be mitigated by smoother interpolation and Gaussian filtering. Notably, reconstructing a 500 × 500 image from a 50 × 50 initialization demonstrates the potential for up to 90% reduction in projection data, corresponding to a similar reduction in radiation dose. These findings highlight the practicality and scalability of the proposed method for quantum-enhanced tomographic reconstruction, offering a promising direction for low-dose, high-fidelity imaging with current-generation quantum devices.

## I. INTRODUCTION

THIS Quantum computing has made rapid progress in recent years. Quantum algorithms can achieve greater efficiency than classical algorithms that use quantum bits (qubits) because it uses all qubits simultaneously by superimposing and/or entanglement [1,2,3]. Parallel computing and qubit decomposition algorithms have also been developed to more efficiently compute quantum algorithms that use many qubits [4,5,6]. In particular, the advent of the D-Wave system's hybrid solver has made it possible to compute quantum optimization algorithms expressed in up to 2 million qubits [7,8,9]. Although the hybrid solver still cannot find a global energy minimum for quantum algorithms representing more than ten thousand qubits, it is sufficiently powerful to test quantum optimization algorithms on practical problems [10,11]. These developments will have a major impact on a wide range of industries and scientific research areas using quantum optimization algorithms.

There are three main types of systems that require tomographic reconstruction algorithms to analyze the internal structure of a sample: spiral computed tomography (CT) [12], electron tomography [13], and synchrotron X-ray tomography [14]. Various classical reconstruction algorithms have been developed to suit each system [15,16,17]. These classical algorithms have become increasingly capable of reconstructing tomographic images more quickly and reliably with advances in hardware. Nevertheless, various types of noise and artifacts generated in each system still create limitations in the accuracy of tomographic images. In particular, motion artifacts were a major cause of corruption of the Helgason-Ludwig consistency conditions of the projection data. To correct this, a virtual alignment method for the pattern of samples in a sinogram has

D-Wave leap quantum cloud service was supported by the Chungbuk Quantum Research Center at Chungbuk National University. H. L. was supported by the National Research Foundation of Korea(NRF) grant funded by the Korea government(MSIT)(RS-2024-00352408). K. J. was supported by the MSIT(Ministry of Science and ICT), Korea, under the ITRC(Information Technology Research Center) support program(IITP-RS-2024-00437284) supervised by the IITP(Institute for Information & Communications Technology Planning & Evaluation). *(Corresponding author: Kyungtaek Jun.)*

H. L. is with Quantum Research Center, QTomo, Chungcheongbuk-do, 28535, South Korea (e-mail: hjlee@qtomo.org).

K. J. is with Quantum Research Center, QTomo, Chungcheongbuk-do, 28535, South Korea and Chungbuk Quantum Research Center, Chungbuk National University, Chungcheongbuk-do, 28644, South Korea (e-mail: ktfriends@chungbuk.ac.kr).

been developed [18,19,20,21], but it is still not easy to apply to existing classical algorithms.

Recently, the development of quantum linear systems [22] has led to the development of quantum optimization tomographic reconstruction algorithms [10]. The quantum algorithm had a problem in finding the global minimum energy when reconstructing around 10,000 qubit tomographic images. A superposition technique on X-ray mass attenuation coefficients for each pixel was developed to find the global minimum energy by minimizing the qubits used and reducing the QUBO matrix coefficient range [11]. This quantum tomographic image reconstruction algorithm was also applicable to real synchrotron radiation projection data, but could only reconstruct up to 100×100 images. Quantum tomographic image reconstruction algorithms have great potential for advancement because they can reconstruct clean CT images than classical CT image reconstruction algorithms [23]. Moreover, quantum algorithms have a great advantage because they can reconstruct tomographic images without errors as long as only 50% of the area within the projection data is ideal [24]. However, the memory issues of the computer used for calculations, the number of qubits that the hybrid solver can compute the global minimum energy, and the energy that can be assigned to couplers and biases are still major obstacles to quantum optimization problems [25].

In this paper, we propose a qubit-efficient iterative reconstruction algorithm for high-resolution tomographic imaging using quantum annealing. The method aims not only to reduce qubit usage but also to significantly reduce the total radiation dose required during projection data acquisition. By initiating the reconstruction from a lower-resolution sinogram and refining it progressively, the total radiation exposure can theoretically be reduced by up to $(1 - n/N) \times 100\%$ where $n$ and $N$ denote the number of pixels along one spatial dimension of the initial and final images, respectively. This concept is formalized in our region-wise optimization framework (see Section 2.2.1). Although the proposed method is demonstrated primarily using parallel-beam geometry for simplicity, it is designed to be compatible with a wide range of tomographic acquisition systems. Since the algorithm operates directly on projection data, it can be applied to any system that uses Charge-Coupled Device (CCD) detectors or other array-based acquisition hardware. By aggregating or pre-processing CCD pixels appropriately, the method is applicable to cone-beam CT, fan-beam CT, synchrotron X-ray tomography, and electron tomography (ET). The region-wise QUBO formulation is agnostic to the specific system geometry, allowing seamless adaptation across modalities.

The method begins by reducing the size of the given sinogram and reconstructing a low-resolution tomographic image using a QUBO-based formulation optimized via a D-Wave hybrid solver. This image is then upscaled to the target resolution, and the upscaled image is divided into multiple spatial regions. Each region is sequentially selected, transformed into a QUBO problem, and refined using quantum optimization, while the rest of the image remains fixed. The refinement process proceeds iteratively, updating the image region by region, enabling high-quality reconstruction with significantly fewer qubits than required in conventional full-image QUBO approaches.

We further investigate an alternative variant of the method by reducing the number of projection angles in the sinogram acquisition step. While the first variant uses a sinogram with 100 projection angles between 0° and 180°, the second variant reconstructs images from a sparsely sampled sinogram generated from only a subset of projection angles. Both variants follow the same region-wise iterative refinement process. Our findings also suggest that this strategy holds strong potential for low-dose tomographic imaging, particularly in clinical and radiation-sensitive experimental environments. Experimental results show that the first method successfully reconstructs a $100 \times 100$ binary tomographic image—partitioned into four $50 \times 50$ regions—with high fidelity, and the second method achieves equally successful reconstruction for a $100 \times 100$ integer-valued image (with pixel values 0–3) under the same four-region iterative refinement strategy.

The main contributions of this work are as follows:

(1) We propose a hybrid quantum-classical framework that performs region-wise QUBO-based reconstruction for tomographic imaging with significantly reduced qubit requirements.

(2) We introduce and compare two sinogram reduction strategies: the first approach reduces the sinogram resolution directly, while the second reduces both the number of projection angles and the resolution of the resulting sparse-view sinogram. Both approaches are followed by the same region-wise iterative refinement pipeline, demonstrating the flexibility and robustness of the proposed framework.

(3) We demonstrate that the proposed method can reconstruct $100 \times 100$ binary and integer-valued tomographic images using only 2,500 and 7,500 qubits, respectively. These numbers are significantly lower than the qubit counts that would be required for full-resolution QUBO formulations, which are known to be impractical for current quantum annealers.

(4) We validate the effectiveness and practicality of the proposed methods using a D-Wave hybrid quantum annealer in both dense-view and sparse-view sinogram scenarios.

(5) By successfully reconstructing tomographic images from sparsely sampled sinograms with reduced projection angles, the proposed method contributes to the potential for low-dose tomographic imaging, which is highly desirable in clinical settings for reducing patient radiation exposure.

(6) The generality of the method allows it to be applied across a variety of tomographic systems and acquisition geometries, as long as appropriate projection data is provided.

The remainder of this paper is organized as follows. Section 2 details the proposed reconstruction framework, including the QUBO formulation, the region-wise refinement procedure, and the two sinogram reduction strategies. Section 3 presents experimental results for both binary and integer-valued tomographic images under full and sparse projection conditions. Section 4 discusses the implications, limitations, and potential future directions of this work.

## II. METHOD

This section presents the proposed iterative tomographic image reconstruction framework using quantum annealing. The method aims to reconstruct high-resolution tomographic images while minimizing qubit requirements by applying region-wise refinement based on QUBO formulations. The overall procedure begins with downscaling the given sinogram, reconstructing a low-resolution image using a D-Wave hybrid solver, and progressively refining upscaled images through localized quantum optimization. We explore two distinct reconstruction strategies based on how the sinogram is reduced: one using full-view sinograms with spatial resizing, and the other using sparse-view sinograms with fewer projection angles. The general framework is first introduced in Section 2.1, followed by detailed descriptions of each reconstruction strategy in Sections 2.2 and 2.3.

We begin by outlining the general reconstruction process based on sinogram downscaling and region-wise refinement.

### A. Overview of the Iterative Quantum Tomographic Image Reconstruction Framework

This section introduces the general framework used in both reconstruction strategies. The proposed method performs high-resolution tomographic image reconstruction by iteratively refining localized image regions using quantum optimization. The input to the method is a sinogram, which may be either dense-view or sparse-view, and it is first resized to a lower resolution. A low-resolution tomographic image is reconstructed from the resized sinogram using a QUBO formulation solved via D-Wave's hybrid quantum solver. This image is then upscaled to the target resolution and partitioned into spatial regions. Each region is refined one by one using partial superposition and quantum optimization, while the rest of the image remains unchanged. The process repeats iteratively across all regions until convergence.

Although the following sections illustrate the method using a parallel-beam geometry for simplicity of explanation, the algorithm is equally applicable to other projection models such as fan-beam, cone-beam, and electron tomography. The iterative quantum tomographic image reconstruction is agnostic to the specific system geometry, making the approach generalizable to a wide class of tomographic reconstruction problems.

We propose two main variants of this method. The first approach uses the full-view sinogram as input and applies single-stage or multi-stage resolution refinement. The second approach uses a sparse-view sinogram constructed from a limited number of projection angles, and applies the same refinement strategy. The following subsections describe each variant in detail.

We begin by describing the full-view sinogram reconstruction strategy.

### B. Full-View Sinogram Reconstruction Strategy
#### 1) Single-Stage Expansion Strategy

Figure 1 illustrates the overall flowchart of the proposed multi-stage expansion strategy. In this section, we focus on the special case of $K = 1$, referred to as the single-stage expansion strategy. Given a sinogram $P \in \mathbb{R}^{N_S \times M_S}$, where $N_S$ is the number of projection angles and $M_S$ is the number of detector bins, we downscale it to a lower-resolution version $\tilde{P} \in \mathbb{R}^{n_S \times m_S}$ by dividing each axis into $d$ segments such that $N_S = n_S \cdot d_1$, $M_S = m_S \cdot d_2$. The resulting downscaled sinogram corresponds to a reconstructed tomographic image of size $n \times n$.

Each $d_1 \times d_2$ patch in the original sinogram is aggregated (e.g., by computing the mean, maximum, or minimum) to form a single pixel in the downscaled sinogram (see Fig. 2a). To avoid introducing artifacts in the central region of the sinogram, the downscaling process aggregates only over pixel groups that do not intersect the rotation axis. In addition to spatial aggregation, downscaling along the projection angle dimension can be performed by either selecting a representative angle (e.g., the central angle within each group of $d_1$ angles), or by computing the mean projection across all $d_1$ angles. This allows the sinogram to be reduced in both spatial and angular resolution while maintaining meaningful structural information. The corresponding tomographic image reconstructed from $P$ is of size $N \times N$, while that from $\tilde{P}$ is $n \times n$. A quantum tomographic reconstruction algorithm is then applied to $\tilde{P}$, resulting in a low-resolution tomographic image of size $n \times n$.

The reconstructed image is then upscaled back to $N \times N$ using interpolation methods such as nearest-neighbor or bilinear interpolation, depending on the reconstruction context. The upscaled image is partitioned into $R_1 = d_1 d_2$ subregions denoted as $\{S_l\}_{l=1}^{R_1}$, where each $S_l \subset \mathbb{R}^2$ is of size $n \times n$. These regions are processed sequentially through the following iterative refinement procedure. Each pixel in region $S_l$ is encoded using $m$ qubits, resulting in $n^2 \cdot m$ qubit variables per region.

(a) A region $S_l$ is selected, and its pixel values within $S_l$ are temporarily set to zero.
(b) The Radon transform is applied to an image in which the selected region $S_l$ has been zeroed out, resulting in a sinogram that reflects the absence of that region's contribution. We refer to this as a zero-masked sinogram $P_z$, as it reflects the projection of a partially masked image.
(c) The difference between the original and zero-masked sinogram is computed to isolate the contribution of region $S_l$.
(d) A superposed sinogram is generated by applying the Radon transform to a hybrid image in which only the region $S_l$ is variable and all other pixels retain their fixed values from the previously reconstructed image. Let $\mathbf{x} \in \{0, 1\}^{n^2 m}$ represent the pixel values in $S_l$, and let $\mathbf{Q} \in \mathbb{R}^{n^2 m \times n^2 m}$ be the QUBO matrix, and $\mathbf{c} \in \mathbb{R}^{n^2 m}$ the linear coefficient. The QUBO formulation is defined as:

$$\min_{\mathbf{x} \in \{0,1\}^{n^2 m}} \mathbf{x}^\top \mathbf{Q} \mathbf{x} + \mathbf{c}^\top \mathbf{x} \qquad (1)$$

The matrix $\mathbf{Q}$ can be represented as an upper triangular matrix for compactness, since the quadratic form remains unchanged. The matrix $\mathbf{Q}$ and the vector $\mathbf{c}$ are derived from the relationship between the superposed sinogram and the difference between the original and zero-masked sinograms, encoding the cost of aligning the qubit-encoded region with the target sinogram contribution. Notably, because each variable $x_i \in \{0,1\}$ is binary, satisfying $x_i^2 = x_i$, the QUBO formulation can equivalently be expressed as $\mathbf{x}^\top \mathbf{Q} \mathbf{x}$ by substituting $x_i x_i$ with $x_i$. In this case, the diagonal elements of matrix $\mathbf{Q}$ are equal to the elements of vector $\mathbf{c}$.

strategy. At each stage $k$, a sinogram of size $n_{k,S} \times m_{k,S}$ (projection angles × detectors) is used to reconstruct a tomographic image of size $n_k \times n_k$. The pipeline is illustrated using parallel-beam CT for clarity, but the same method applies to cone-beam, fan-beam, and other tomographic systems.

(e) The QUBO is solved using the D-Wave hybrid solver to update the values within region $S_l$.
(f) The updated image is passed to the next iteration to refine the subsequent region.

The complete workflow of the single-stage expansion and region-wise refinement strategy is summarized in Fig. 1, providing a visual overview of the iterative reconstruction process.

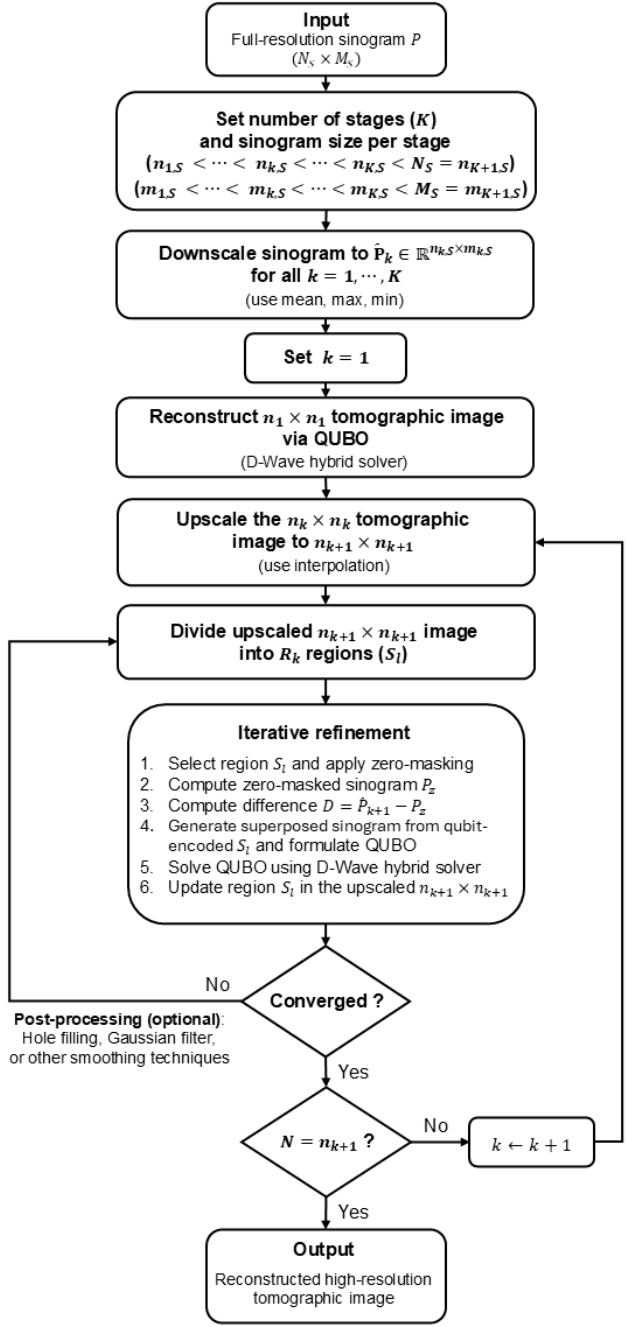

Fig. 1. Flowchart of the proposed multi-stage expansion strategy for tomographic image reconstruction, where the process is repeated over $K$ stages. The case of $K = 1$ corresponds to the single-stage expansion

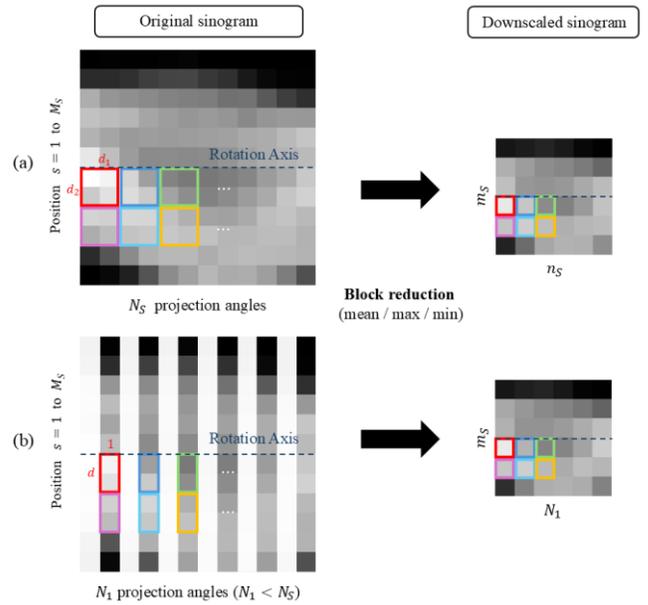

Fig. 2. Downscaling strategies for sinogram size reduction. (a) Full-view sinogram of size $N_S \times M_S$ is reduced to $n_S \times m_S$ by applying block reducing over each $d_1 \times d_2$ patch, where $N_S = n_S \cdot d_1, M_S = m_S \cdot d_2$. (b) Sparse-view sinogram of size $N_1 \times M_S$ with fewer projection angles ($N_1 < N_S$) is reduced to $N_1 \times m_S$ by applying block reducing over each $1 \times d$ vertical patch along the position axis in the projection image.

This process is repeated for all $R_1$ regions. If convergence is not achieved, post-processing techniques such as hole filling based on pixel connectivity or filtering with Gaussian kernels may be applied to improve local continuity before repeating the refinement loop. While the image is typically divided into non-overlapping regions to reduce computationalf cost and ensure coverage, the method does not require strict non-overlapping partitioning. In practice, regions may overlap, or selected areas can be re-optimized multiple times based on specific reconstruction goals or prior knowledge.

To represent the pixel values in region $S_l$ as qubit variables, each pixel $I_{i,j} \in S_l$ is encoded using $m$ binary variables as follows [10]:

$$I_{i,j} \approx \sum_{k=0}^{m-1} 2^k q_k^{i,j}, \quad q_k^{i,j} \in \{0,1\}. \tag{2}$$

This binary formulation follows the radix-2 representation, where each integer pixel value is approximated using a weighted sum of binary variables. This approach is widely used in QUBO modeling to efficiently encode multi-bit discrete variables using binary qubits. Here, $m$ is determined by the dynamic range of the image: for binary images, $m = 1$; for integer-valued images, $m$ is chosen such that $2^m$ exceeds the maximum pixel intensity. This formulation allows each pixel value to be represented using a compact binary code and is compatible with standard QUBO-based optimization.

Alternatively, if the X-ray mass attenuation coefficients (MACs) of the sample are known, the pixel values can be expressed as a weighted sum of predefined MAC values [11]. Let $\alpha_k = \mu_k/\sigma_k$ denote the $k$-th MAC value for the material, where $m$ such values are known. Then each pixel can be represented as:

$$I_{i,j} = \sum_{k=1}^{m} \alpha_k q_k^{i,j}, \quad q_k^{i,j} \in \{0,1\}, \tag{3}$$

such that $I_{i,j} \in \{0, \alpha_1, \alpha_2, \cdots, \alpha_m\}$. To ensure monotonicity and improve stability in optimization, an alternative cumulative encoding can be applied:

$$J_{i,j} = \alpha_1 + \sum_{k=2}^{m}(\alpha_k - \alpha_{k-1}) q_k^{i,j} \tag{4}$$

This representation ensures that $J_{i,j} \in [\alpha_1, \alpha_m]$ and is particularly suitable when prior physical knowledge of the sample's composition is available. The choice between binary encoding and MAC-based encoding depends on the reconstruction context and the availability of prior information.

When applying the Radon transform to the qubit-encoded image, each pixel in the resulting superposed sinogram can be expressed as a linear combination of the encoded pixel values. Specifically, for a given projection angle $\theta$ and detector position $s$, the superposed sinogram value $IP(\theta, s)$ is computed as:

$$IP(\theta, s) = \sum_{i,j} c_{ij} I'_{i,j} \tag{5}$$

where $I_{i,j}'$ is the encoded pixel value at position $(i, j)$, and $c_{ij}$ denotes the projection weight defined by the Radon transform geometry. The set of weights $\{c_{ij}\}$ captures how each image pixel contributes to the projection ray defined by angle $\theta$ and position $s$.

The QUBO objective for region $S_l$ is then formulated using the difference between the original and zero-masked sinograms as follows:

$$\sum_{(\theta,s)} [\{IP - (P - P_z)\}(\theta, s)]^2 \tag{6}$$

where $(\theta, s)$ denotes each pixel position in the sinogram of size $N_S \times M_S$, $P(\theta, s)$ is the original sinogram, $P_z(\theta, s)$ is the zero-masked sinogram, and $IP(\theta, s)$ is the superposed sinogram generated by applying the Radon transform to an image where only region $S_l$ is qubit-encoded.

We let

$$D(\theta, s) = P(\theta, s) - P_z(\theta, s) \tag{7}$$

Denote the contribution of region $S_l$ to the sinogram. The QUBO objective is then written as:

$$\sum_{(\theta,s)} \{IP(\theta, s) - D(\theta, s)\}^2 \tag{8}$$

$$= \sum_{(\theta,s)} \left\{ \sum_{i,j} c_{ij} I'_{i,j} - D(\theta, s) \right\}^2 \tag{9}$$

$$= \sum_{(\theta,s)} \left[ \left( \sum_{i,j} c_{ij} I'_{i,j} \right)^2 - 2D(\theta, s) \left( \sum_{i,j} c_{ij} I'_{ij} \right) + \{D(\theta, s)\}^2 \right] \tag{10}$$

By moving the constant term $\sum_{(\theta,s)} \{D(\theta, s)\}^2$ to the left-hand side, the objective can be expressed in the standard QUBO form:

$$\mathbf{x}^\top \mathbf{Q} \mathbf{x} + \mathbf{c}^\top \mathbf{x}$$

where $\mathbf{x} \in \{0, 1\}^{n^2 m}$ represents the qubit-encoded pixel values in region $S_l$. The minimum value of this QUBO is equal to $-\sum_{(\theta,s)} \{D(\theta, s)\}^2$.

### 2) Multi-Stage Expansion Strategy

As illustrated in Fig. 1, the proposed multi-stage expansion strategy incrementally reconstructs a high-resolution tomographic image of size $N \times N$ through a sequence of $K$ upscaling stages. The process begins by setting the total number of stages $K$, and determining the intermediate image resolutions $\{n_k\}_{k=1}^{K}$ such that $n_1 < n_2 < \cdots < n_K < N$. For each stage $k$, a corresponding sinogram resolution $n_{k,S} \times m_{k,S}$ is chosen, satisfying $n_{k,S} < N_S$ and $m_{k,S} < M_S$, where $N_S \times M_S$ is the resolution of the full input sinogram $P$. The original sinogram $P \in \mathbb{R}^{N_S \times M_S}$ is then downscaled to $\tilde{P}_k \in \mathbb{R}^{n_{k,S} \times m_{k,S}}$ by aggregating non-overlapping $d_1 \times d_2$ patches, where aggregation is typically performed using operations such as mean, minimum, or maximum.

At the first stage ($k = 1$), the low-resolution sinogram $\tilde{P}_1$ is used to formulate a QUBO problem, which is solved using D-Wave's hybrid solver. This produces a reconstructed image of size $n_1 \times n_1$. The resulting image is then upscaled to $n_2 \times n_2$ using an interpolation method, such as nearest-neighbor, bilinear, or bicubic interpolation, depending on the reconstruction context. The upscaled image is partitioned into $R_k$ subregions $\{S_l\}_{l=1}^{R_k}$, each of size $n_1 \times n_1$. For each region $S_l$, we apply the localized refinement procedure described in Section 2.2.1, which includes Steps (a) through (f). After all subregions have been reconstructed, we obtain an updated image of size $n_2 \times n_2$, which is used to generate a sinogram. This sinogram is then compared to the precomputed downscaled sinogram $\tilde{P}_2$.

If the generated sinogram sufficiently matches $\tilde{P}_2$, we increment $k$ by one and repeat the process: upscale the image to $n_{k+1} \times n_{k+1}$, divide it into subregions, and perform refinement. If convergence is not achieved, post-processing techniques such as hole filling, Gaussian filtering, or other smoothing methods may be applied before re-entering the refinement loop for the current stage.

This iterative upscaling and refinement process continues until the generated sinogram at stage $k$ converges to $\tilde{P}_{k+1}$, and the final image resolution reaches $N \times N$. At this point, the process terminates, yielding the final high-resolution reconstructed image.

### C. Sparse-View Sinogram Reconstruction Strategy

In this variant, the input sinogram is generated from a reduced number of projection angles (e.g., $N_1 < N_S$ ). Although uniform angular spacing generally yields better reconstruction quality, the proposed refinement framework does not require strict angular uniformity and can be applied to sinograms with non-uniform or irregular sampling patterns. Each projection image is then resized along the spatial dimension to reduce its resolution to $m_S$. The resulting sinogram has dimensions $N_1 \times m_S$. Downscaling can be performed using the mean, maximum, or minimum across grouped pixels, as illustrated in Fig. 2b.

Despite starting from a sparse-view sinogram, the same reconstruction pipeline is applied:
(i) Quantum reconstruction from the downscaled sparse-view sinogram to produce a low-resolution image.
(ii) Upscaling to the target resolution.
(iii) Region-wise iterative refinement using the same partial superposition and QUBO optimization strategy described in Section 2.2.

This strategy directly builds upon the framework introduced in the previous sections. Despite the sinogram being sparse in angular views, both the single-stage and multi-stage expansion strategies described earlier can be applied without modification. In the single-stage case, the low-resolution image is reconstructed from the sparse-view sinogram and then upscaled to the final resolution. In the multi-stage scenario, intermediate resolutions can be introduced, and region-wise refinement is performed at each stage following the same process: quantum reconstruction, interpolation-based upscaling, region partitioning, and localized QUBO optimization. The ability to reuse the full reconstruction pipeline ensures that the proposed method remains effective even under angular undersampling, offering strong potential for low-dose CT applications.

The effectiveness of the proposed framework is validated through experiments on both full-view and sparse-view sinograms, demonstrating its ability to achieve high-resolution reconstruction with reduced qubit usage and lower projection requirements.

## III. RESULT

To assess the effectiveness of the proposed method, we present experimental results for both full-view and sparse-view sinograms using binary and integer-valued phantoms.

### A. Experiment on Binary Shepp-Logan Phantom

To evaluate the proposed method under idealized conditions, we conducted an experiment using a binary Shepp-Logan phantom. Following the single-stage expansion strategy described in Section 2.2.1, the parameters were set to $N = 100, d_1 = d_2 = 2$, and $n = 50$, resulting in four $50 \times 50$ regions for iterative refinement. Since the phantom is binary, each pixel was encoded using a single qubit ($m = 1$). A sinogram of size $100 \times 100$ was generated by applying the Radon transform to the $100 \times 100$ binary Shepp-Logan image using 100 projection angles. To reduce its resolution, average pooling with a $2 \times 2$ window was applied via 'block_reduce' from scikit-image, resulting in a $50 \times 50$ downscaled sinogram. To reduce the number of projection angles, the original 100 uniformly sampled angles from 0° to 180° were grouped into non-overlapping pairs within each $2 \times 2$ patch. For each pair, the angle corresponding to the second projection (i.e., the higher angle) was selected, resulting in 50 projection angles uniformly spaced from 0° to 180°. This effectively downsampled the angular resolution while preserving uniform angular coverage. Although the downscaled sinogram maintained a similar range of pixel values to the original $100 \times 100$ sinogram, a correction was required due to the geometric properties of the Radon transform. Since the Radon transform represents line integrals through the object, reducing the image resolution by half effectively shortens the X-ray path length through the object by a factor of 2. To account for this change, all values in the downscaled sinogram were multiplied by $\frac{1}{2}$ to normalize the line integrals with respect to the reduced object size.

A QUBO formulation was constructed based on the reduced sinogram, yielding a known target minimum energy of $-1,533,290.16$. The minimum energy obtained by the D-Wave hybrid solver was $-1,532,543.73$. This $50 \times 50$ image was then upscaled to $100 \times 100$ using nearest-neighbor interpolation. The resulting image is shown as the *initial upscaled image* in the first row of Fig. 3. The upscaled image was divided into four non-overlapping $50 \times 50$ regions, and the region-wise refinement process described in Section 2.2.1 was applied sequentially to each. As seen in the "Iteration 1" row of Fig. 3, applying the method once to each region noticeably sharpened edges in the reconstructed image. However, residual dot artifacts remained in the interior of the object. Repeating the region-wise refinement process once more over all four regions eliminated these artifacts and yielded a final reconstruction identical to the original tomographic image (see "Iteration 2" row in Fig. 3).

Table 1 provides quantitative insight into the iterative refinement process. From Iteration 1 to Iteration 2, the difference between the target minimum and the minimum energy obtained from the D-Wave solver progressively decreased, indicating improved optimization. Notably, for the final region in Iteration 2, this difference was reduced to as low as $1.5 \times 10^{-7}$, demonstrating a successful and nearly perfect reconstruction.

TABLE I
DETAILED QUANTUM OPTIMIZATION RESULTS FROM THE D-WAVE HYBRID SOLVER FOR EACH REGION AND ITERATION IN BOTH RECONSTRUCTION EXAMPLES. EACH ROW CORRESPONDS TO A SPECIFIC REGION $S_l$ DURING A PARTICULAR ITERATION APPLIED TO EITHER THE BINARY (100 × 100) OR INTEGER-VALUED (100×100, PIXEL VALUES ∈ {0,1,2,3}) SHEPP-LOGAN PHANTOM. THE TABLE REPORTS THE SOLVER RUNTIME IN SECONDS, THE THEORETICAL TARGET MINIMUM ENERGY COMPUTED FROM THE QUBO FORMULATION, THE MINIMUM ENERGY OBTAINED BY D-WAVE, AND THE ABSOLUTE DIFFERENCE BETWEEN THE TWO.

| SAMPLE | ITERATION | REGION | Run time (s) | Target minimum | Minimum energy by D-Wave | Absolute difference between two |
|---|---|---|---|---|---|---|
| Binary image (100 × 100) | 1 | $S_1$ | 6.35 | -2865018.9247 | -2813456.9991 | 51561.9256 |
| | | $S_2$ | 6.36 | -3471534.2524 | -3433782.4013 | 37751.8513 |
| | | $S_3$ | 6.36 | -2888022.8897 | -2864521.3878 | 23501.5020 |
| | | $S_4$ | 6.36 | -3543823.1399 | -3542119.4029 | 1703.7370 |
| | 2 | $S_1$ | 6.36 | -2881565.1607 | -2880533.8795 | 1031.2812 |
| | | $S_2$ | 6.32 | -3473412.6393 | -3472611.9267 | 800.7126 |
| | | $S_3$ | 6.35 | -2912525.4048 | -2912380.6420 | 144.7628 |
| | | $S_4$ | 6.36 | -3498681.9010 | -3498681.9010 | 1.5E-07 |
| Integer valued image (100 × 100, values: 0-3) | 1 | $S_1$ | 27.30 | -2145325.7413 | -2087908.8500 | 57416.8913 |
| | | $S_2$ | 27.27 | -2458946.5805 | -2421684.3338 | 37262.2467 |
| | | $S_3$ | 27.25 | -1694428.5912 | -1669147.5329 | 25281.0582 |
| | | $S_4$ | 27.27 | -2075670.9377 | -2074236.1337 | 1434.8041 |
| | 2 | $S_1$ | 27.22 | -2125784.8366 | -2110864.8883 | 14919.9483 |
| | | $S_2$ | 27.28 | -2466651.1653 | -2457976.5742 | 8674.5911 |
| | | $S_3$ | 27.23 | -1718175.9469 | -1714074.5188 | 4101.4281 |
| | | $S_4$ | 27.25 | -2033356.8569 | -2033356.8569 | 3.2E-07 |

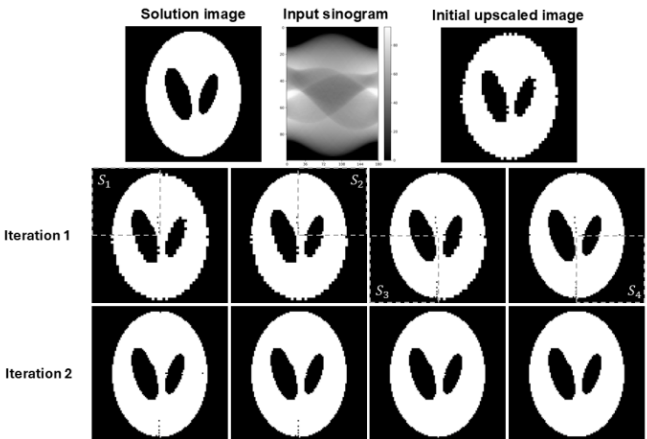

Fig. 3. Iterative tomographic reconstruction results for the binary Shepp-Logan phantom using the proposed method (Section 2.2.1). The first row shows (from left to right): the target 100 × 100 binary Shepp-Logan image, the input full-resolution sinogram (100 × 100), and the upscaled 100 × 100 image from the initial reconstructed 50 × 50 image obtained via QUBO, prior to refinement. The second row presents the results after the first refinement iteration, where each image corresponds to the update of regions $S_1, S_2, S_3,$ and $S_4$, respectively. Region boundaries are highlighted in these images. The third row shows the results after the second refinement iteration, completing one full cycle across all regions. "Iteration 1" and "Iteration 2" labels indicate the corresponding stages of refinement.

### B. Experiment on Sparse-View Sinogram with Integer-Valued Phantom

To evaluate the effectiveness of the proposed method on sparse-view data, we generated a 100 × 100 Shepp-Logan sample image in which each pixel takes an integer value from 0 to 3. The original sinogram was created by applying the Radon transform using 50 uniformly spaced projection angles over 180 degrees, resulting in a 50 × 100 sinogram. Following the downscaling strategy described in Section 2.3, we fixed the angular resolution and applied average pooling along the detector axis using a 1 × 2 window, reducing the sinogram to a size of 50 × 50. Because the target image size was reduced by a factor of two, and the Radon transform represents line integrals through the image, all values in the downscaled sinogram were scaled by $\frac{1}{2}$ to account for the reduced path length. The reconstruction parameters were set to $N = 100, N_1 = 50, d = 2$, and $n = 50$.

Each pixel in the tomographic image was represented using the encoding method in (4):

$$I_{i,j} = q_1^{i,j} + q_2^{i,j} + q_3^{i,j}, \quad q_k^{i,j} \in \{0, 1\},$$

allowing for values from 0 to 3 using three binary variables. Using the reduced sinogram, a QUBO formulation was constructed, resulting in a target minimum value of −1,952,866.74. The minimum energy obtained from the D-Wave hybrid solver was −1,951,405.24. The reconstructed 50 × 50 image appears in the first row of Fig. 4 under the label *Reconstructed image*. The image was then upscaled to 100 × 100 using interpolation and divided into four non-overlapping 50 × 50 regions. The region-wise refinement process (steps (a)–(f) in Section 2.2.1) was applied sequentially to each region. As shown in Fig. 4 ("Iteration 1" row), the result after the first iteration included multiple colored dot artifacts within the object. These discontinuities limited the effectiveness of local refinement steps and prevented significant error reduction, particularly in the lower two regions. To address this issue, interpolation-based upscaling was used instead, followed by Gaussian filtering with $\sigma = 1$, which suppressed the dot artifacts and enabled the refinement process to converge accurately in the second iteration. After applying Gaussian filtering with $\sigma = 1$, the second refinement iteration successfully eliminated the artifacts, resulting in a final image that closely matched the original phantom (see "Iteration 2" row in Fig. 4).

Table 1 also provides insight into the behavior of the iterative refinement process in the second experiment. During Iteration 1, the difference between the target minimum and the D-Wave-obtained minimum energy gradually decreased across all regions. However, after applying Gaussian filtering before Iteration 2, the difference for region $S_1$ increased by nearly an order of magnitude compared to the previous iteration. This behavior is expected since the Gaussian filter smooths the image, temporarily disturbing the previously optimized structures and increasing the energy gap for the first region in Iteration 2. Despite this, the difference steadily decreased again for the remaining regions in Iteration 2, reaching as low as $3.2 \times 10^{-7}$ in the final step. This result suggests that the artifacts introduced by filtering were effectively removed through the iterative refinement process.

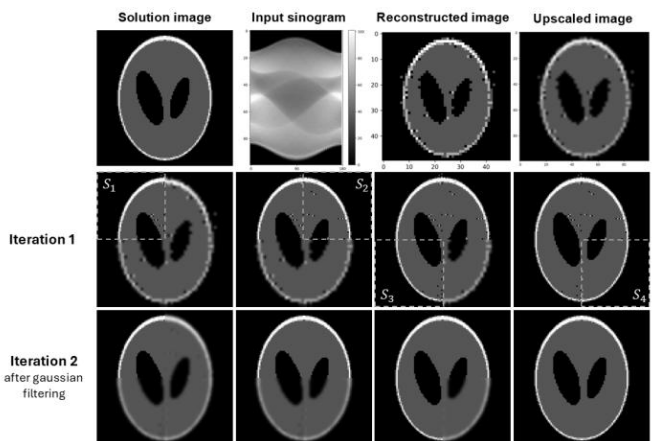

Fig. 4. Iterative tomographic reconstruction results for the integer-valued Shepp-Logan phantom using the sparse-view strategy (Section 2.3). The first row shows (from left to right): the target 100 × 100 phantom image with pixel values in $[0, 3]$, the input full-resolution sinogram (50 × 100), the initial reconstructed 50 × 50 image, and the upscaled 100 × 100 image prior to refinement. The second row presents the results after the first region-wise refinement iteration (Iteration 1). Gaussian filtering ($\sigma = 1$) was then applied as post-processing, and the third row shows the results of the second refinement iteration that followed.

## IV. Discussion

The proposed quantum-assisted tomographic image refinement method achieved accurate image recovery with significantly reduced qubit requirements by combining sinogram downscaling and localized iterative refinement. Beyond the empirical performance observed in Sections 3.1 and 3.2, several aspects of the method's flexibility and behavior warrant further discussion and suggest directions for future development.

First, although our implementation used non-overlapping $n \times n$ regions for refinement, the framework is inherently flexible. Overlapping patches may be introduced to mitigate discontinuities across regional boundaries and improve continuity in the reconstructed image. As seen in Fig. 3, dot-like artifacts were occasionally observed near region boundaries after a single iteration. Including these transition areas within overlapping refinement windows may allow for more accurate correction. Furthermore, the framework supports selective reprocessing of arbitrary regions beyond the original segmentation, enabling adaptive refinement based on prior knowledge or intermediate results. Furthermore, rather than reapplying refinement uniformly across all regions, the framework allows individual regions to be revisited selectively based on intermediate reconstruction results. After a few iterations, if specific areas still exhibit artifacts or insufficient detail, those regions can be isolated, zero-masked, and refined again using the same quantum-assisted procedure. This adaptive targeting is more efficient and can be particularly valuable in reducing computational cost while maintaining high reconstruction fidelity where needed. This flexibility is particularly useful in clinical settings, where certain anatomical structures may require higher reconstruction fidelity than others.

Second, our experiments with integer-valued phantoms revealed that the choice of upsampling method significantly influences convergence behavior. Using nearest-neighbor interpolation introduced sharp discontinuities along edges, which hindered the ability of the iterative process to reduce error—especially in the lower image regions—likely due to entrapment in local minimum. Switching to interpolation-based upsampling reduced boundary artifacts, and applying Gaussian filtering with $\sigma = 1$ before the second iteration further smoothed localized noise. This adjustment enabled the refinement loop to converge successfully and highlights the importance of initialization smoothness in quantum-assisted reconstruction tasks.

In addition to the above considerations, it is important to discuss alternative quantum strategies and the limitations imposed by current hardware. One such approach is what we refer to as *simple decomposition*, serves as a conceptual baseline. In this method, a QUBO model is constructed by comparing the predicted and measured projection values. Specially, the error term minimized is $\big(IP(\theta, s) - P(\theta, s)\big)^2$, where $P(\theta, s)$ denotes the measured sinogram value, and $IP(\theta, s)$ is the projection value derived from the current qubit-encoded image estimate, where each pixel is represented using binary variables within the QUBO formulation. This corresponds to the case where the projection segment length $w = 1$. For $w > 1$, the model extends this objective to a group of $w$ adjacent detector positions within a single projection angle, resulting in a sum of squared differences over the window. In both cases, only the tomographic image pixels that contribute to the selected projection segment are updated during optimization, while the rest remain fixed. This process is then repeated iteratively for all segments across all projection angles.

The total number of qubits required is the sum over all regions of the number of contributing image pixels multiplied by the number of qubits per pixel. If this total remains below the qubit limit of the quantum system, the entire image can theoretically be reconstructed in a piecewise manner. For an $N \times N$ image, the number of variables influenced by a single projection line $IP(\theta, s)$ is generally less than $3N$ (derived from $(w + 2)N$ with $w = 1$). This implies that, even when using the QUBO refinement [6], the maximum reconstructible image size under a 10,000-qubit constraint is approximately $1700 \times 1700$.

In cases where the image contains three distinct X-ray mass attenuation coefficients, such as in a $500 \times 500$ phantom, reconstruction using the MAC-aware QUBO model [11] is feasible up to around $1000 \times 1000$ resolution. However, even for a single iteration of the QUBO model based on minimizing $\bigl(IP(\theta,s) - P(\theta,s)\bigr)^2$, rough estimates indicate that approximately $O(10^6)$ annealing computations are required. As a result, the method proposed in this paper emphasizes *region-wise refinement* rather than full-image decomposition. In particular, the proposed approach shows greater computational efficiency and convergence robustness by operating on spatial subregions, while also enabling structured prior incorporation and control over qubit allocation.

When reconstructing under known mass attenuation coefficients (as in the experiments of this paper), it is preferable to eliminate the target region via zero-padding and apply the full quantum refinement process directly—especially when using the MAC-aware formulation in (3) or (4). Conversely, when using radix-2 representation as in (2), reconstruction by progressively updating individual pixel values may offer advantages. Looking ahead, we anticipate that hybrid strategies combining region-wise refinement with local simple decomposition inside subregions may further improve scalability and qubit efficiency for high-resolution tomographic reconstruction.

Finally, the choice of the initial region size $n$ plays a crucial role in balancing reconstruction quality and computational feasibility. If $n$ is set too small relative to the target resolution $N$, the low-resolution reconstruction may suffer from excessive noise or structural degradation, reducing the effectiveness of subsequent refinement. While the experiments used $n = 50$ for $N = 100$, further investigation is needed to determine how small $n$ can be set while enabling stable convergence and accurate reconstruction. Additionally, we anticipate that the refinement process may converge more rapidly when the initial $n \times n$ image retains key structural features of the target $N \times N$ tomographic image. If the low-resolution sinogram sufficiently encodes the dominant features, the subsequent upscaling and refinement steps are more likely to preserve and enhance these features efficiently.

Taken together, these findings demonstrate that the proposed framework is not only effective but also adaptable. It allows flexible control over spatial resolution, reconstruction fidelity, and resource constraints, making it a promising approach for quantum-enhanced CT image reconstruction—particularly in low-dose or sparse-view acquisition scenarios. For instance, if a high-resolution $500 \times 500$ tomographic image is reconstructed starting from a $50 \times 50$ initialization, the proposed method can reduce the required projection data—and hence the associated radiation dose—by approximately $90\%$. This highlights the method's potential for low-dose imaging applications, especially in clinical and synchrotron-based systems where minimizing exposure is critical. Future research will further explore its application to real-world medical images, integration with classical postprocessing techniques, and optimization under evolving quantum hardware capabilities.

Although the experiments focused on parallel-beam CT geometry, the proposed framework is compatible with a variety of tomographic acquisition systems. The QUBO-based refinement operates at the level of projection-image relationships, making it adaptable to different system geometries including cone-beam CT, fan-beam CT, synchrotron tomography, and ET. This generality makes it a promising candidate for broader deployment across diverse imaging modalities.

## APPENDIX AND THE USE OF SUPPLEMENTAL FILES

.

## ACKNOWLEDGMENT

## REFERENCES

[1] Mitarai, K., Negoro, M., Kitagawa, M., & Fujii, K. (2018). Quantum circuit learning. Physical Review A, 98(3), 032309.
[2] Wurtz, J., & Love, P. J. (2021). Classically optimal variational quantum algorithms. IEEE Transactions on Quantum Engineering, 2, 1-7.
[3] Arute, F., Arya, K., Babbush, R., Bacon, D., Bardin, J. C., Barends, R., ... & Martinis, J. M. (2019). Quantum supremacy using a programmable superconducting processor. Nature, 574(7779), 505-510.
[4] Pelofske, E., Hahn, G., & Djidjev, H. N. (2022). Parallel quantum annealing. Scientific Reports, 12(1), 4499.
[5] Lee, H., & Jun, K. (2025). Range dependent Hamiltonian algorithms for numerical QUBO formulation. Scientific Reports, 15(1), 8819.
[6] Lee, H., & Jun, K. (2024). QUBO Refinement: Achieving Superior Precision through Iterative Quantum Formulation with Limited Qubits. arXiv preprint arXiv:2411.16138.
[7] O'Malley, D. & Vesselinov, V. V. ToQ.jl: A high-level programming language for D-Wave machines based on Julia, in 2016 IEEE High Performance Extreme Computing Conference (HPEC) (2016) pp. 1–7 https://doi.org/10.1109/HPEC.2016.7761616
[8] McGeoch, C., Farre, P. & Bernoudy, W. D-Wave hybrid solver service+ advantage: Technology update (Tech Rep, 2020).
[9] " D-wave ocean software documentation," https://docs.ocean.dwavesys.com, accessed: 2024-12-03
[10] Jun, K. (2023). A highly accurate quantum optimization algorithm for CT image reconstruction based on sinogram patterns. Scientific Reports, 13(1), 14407.
[11] Jun, K. (2023). Quantum optimization algorithms for CT image segmentation from X-ray data. arXiv preprint arXiv:2306.05522.
[12] Jacops, R., Mraiwa, N., Steenberghe, D., Gijbels, F. & Quirynen, M. Appearance, locationi course and mor- fology of the mandibular incisive canal on spiral CT scan. Dentomaxillofac. Radiol. 31, 322–327 (2002).
[13] Hoff, J. A. et al. Age and gender distributions of coronary artery calcium detected by electron beam tomography in 35,246 adults. Am. J. Cardiol. 87, 1335–1339 (2001).
[14] Böhm, T. et al. Quantitative synchrotron X-ray tomography of the material-tissue interface in rat cortex implanted with neural probes. Sci. Rep. 9, 7646 (2019).
[15] Wang, X. et al. High performance model based image reconstruction. ACM SIGPLAN Not. 51(8), 1–12 (2016).
[16] Schomberg, H. & Timmer, J. The gridding method for image reconstruction by Fourier transformation. IEEE Trans. Med. Imaging 14(3), 596–607 (1995).
[17] Singh, R. et al. Artificial intelligence in image reconstruction: The change is here. Phys. Med. 79, 113–125 (2020).
[18] Jun, K. and Yoon, S. "Alignment Solution for CT Image Reconstruction using Fixed Point and Virtual Rotation Axis", Sci. Rep. 7, 41218 (2017).
[19] Jun, K. and Kim, D. "Alignment theory of parallel-beam computed tomography image reconstruction for elastic-type objects using virtual focusing method", Plos One, 13(6), e0198259 (2018).
[20] Jun, K. and Jung, J. "Virtual multi-alignment theory of parallel-beam CT image reconstruction for elastic objects", Sci. Rep. 9 (2019) 6847.
[21] Jun, K. "Virtual multi-alignment theory of parallel-beam CT image reconstruction for rigid objects", Sci. Rep. 9 (2019) 13518.
[22] Jun, K. (2024). QUBO formulations for a system of linear equations. Results in Control and Optimization, 14, 100380.

[23] Dremel, K., Prjamkov, D., Firsching, M., Weule, M., Lang, T., Papadaki, A., ... & Fuchs, T. O. (2025). Utilizing Quantum Annealing in Computed Tomography Image Reconstruction. IEEE Transactions on Quantum Engineering.

[24] Lee, H., & Jun, K. (2025). Quantum Supremacy in Tomographic Imaging: Advances in Quantum Tomography Algorithms. arXiv preprint arXiv:2502.04830.

[25] Jun, K., & Lee, H. (2023). HUBO and QUBO models for prime factorization. Scientific Reports, 13(1), 10080.